\documentclass[conference]{IEEEtran}
\IEEEoverridecommandlockouts
\usepackage{cite}
\usepackage{amsmath,amssymb,amsfonts}
\usepackage{algorithmic}
\usepackage{graphicx}
\usepackage{textcomp}
\usepackage{xcolor}
\usepackage{afterpage}
\def\BibTeX{{\rm B\kern-.05em{\sc i\kern-.025em b}\kern-.08em
    T\kern-.1667em\lower.7ex\hbox{E}\kern-.125emX}}
\begin{document}

\title{LifeSync-Games: Toward a Video Game Paradigm for Promoting Responsible Gaming and Human Development\\
}

\author{\IEEEauthorblockN{Roberto González-Ibáñez}
\IEEEauthorblockA{\textit{Departamento de Ingeniería Informática}\\
\textit{InTeractiOn Research Lab}\\
\textit{Universidad de Santiago de Chile}\\
\textit{Santiago, Chile}\\
roberto.gonzalez.i@usach.cl}
\and
\IEEEauthorblockN{Joaquín Ignacio Macías-Cáceres}
\IEEEauthorblockA{\textit{Departamento de Ingeniería Informática}\\
\textit{InTeractiOn Research Lab}\\
\textit{Universidad de Santiago de Chile}\\
\textit{Santiago, Chile}\\
joaquin.macias@usach.cl}
\and
\IEEEauthorblockN{Marco Villalta Paucar}
\IEEEauthorblockA{\textit{Escuela de Psicología}\\
\textit{Universidad de Santiago de Chile}\\
\textit{Santiago, Chile}\\
marco.villalta@usach.cl}
}

\maketitle

\begin{abstract}
Technological advancements have made video games a central part of the digital lives of nearly 3 billion people worldwide. Although games can address \mbox{various} social, physical, and psychological needs, their potential to support human \mbox{development} and well-being remains underutilized. Research highlights both negative effects — such as addiction and isolation — and positive outcomes like cognitive improvements and problem-solving skills. However, public discourse and regulation often focus more on risks than benefits. To address this imbalance, we present LifeSync-Games, a framework leveraging simplified digital twins to connect virtual gameplay with real-life activities. This reciprocal relationship aims to enhance the \mbox{developmental} value of gaming by promoting self-regulation and fostering growth across physical, mental, and social domains. We present the framework’s theoretical foundations, technological components, design guidelines, and evaluation approaches. Additionally, we present early applications in both new and bestselling games to demonstrate its versatility and practical relevance.
\end{abstract}



\section{Introduction}


The technological shift of the 21st century has made video games a key part of digital life, with nearly 3 billion active players in 2022 \cite{Palma-Ruiz2022-GamingIndustry}. While these virtual \mbox{environments} address \mbox{various} human and social needs, their potential to support human \mbox{development} remains largely untapped.

Research in psychology and media \mbox{studies} has identified both negative (e.g., addiction, inactivity, isolation \cite{DeLisi2013-ViolentVideoGames}) and positive effects (e.g., cognitive gains, problem-solving \cite{Johannes2021-VideogameWellbeing}) of video games. However, public debate and regulation often emphasize risks over benefits, except in cases involving serious games \cite{Damaševičius2023-SeriousGamesGamification}.

In this context, we introduce LifeSync-Games (formerly blended-Games or bGames), a framework that applies digital twin concepts to create reciprocal links between real-life \mbox{\mbox{activities}} and virtual gameplay. Its goal is to foster self-regulated play while supporting social, physical, and mental well-being. We outline the framework and its exploratory applications, aiming to enhance the positive impact of video games and mitigate their potential harms.

\subsection{Problem statement and justification}

The real and virtual worlds often remain disconnected, with an asymmetric relationship: while video games can impact real life — positively or negatively — real-world actions seldom influence gameplay. This divide is also evident in player motivations, which range from extrinsic factors (e.g., rewards and recognition) to intrinsic ones (e.g., personal achievement and in-game progress) \cite{Ryan2000-IntrinsicAndExtrinsicMotiv} (Figure 1).

\begin{figure}[htbp]
\centerline{\includegraphics[angle=0, width=9cm]{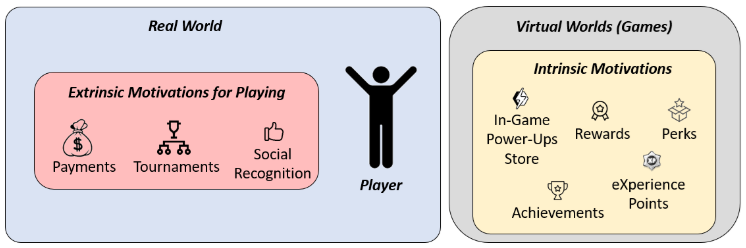}}
\caption{Model of separate worlds with motivations that encourage the use of video games.}
\label{fig1-sccc}
\end{figure}

While video games can offer cognitive and emotional benefits \cite{Johannes2021-VideogameWellbeing}, \mbox{excessive} use is linked to sedentary \mbox{behavior} and health risks, including addiction and, in rare cases, fatal outcomes \cite{DeLisi2013-ViolentVideoGames}. Many players turn to games as an escape, often projecting idealized selves through avatars \cite{Rehak2013-PsychoanalysisAvatar}. However, this identification may negatively impact real-world identity, as self-perception can be temporarily shaped by the virtual character’s success or failure.


\section{Related Work}


Research on video games covers cultural, gameplay, and practical applications across \mbox{various} fields. Game \mbox{studies} — spanning both analog and digital formats — offer theoretical and methodological tools to analyze games, players, and their broader contexts, often emphasizing qualitative \mbox{approaches} \cite{Palaus2017-NeuralBasicsOfVG}.

Four concepts that are often confused with LifeSync-Games can be distinguished as follows:

\begin{itemize}
    \item Serious games: Designed for purposes beyond pure \mbox{entertainment}, such as promoting physical well-being, \mbox{\mbox{addressing}} health conditions, or supporting education \cite{Damaševičius2023-SeriousGamesGamification}. In this type of games, the player's real-world actions do not directly influence gameplay, so the connection between the real and virtual worlds is not reciprocal.
    \item Affective games: These utilize the player’s physiological signals, captured by sensors, to influence gameplay, but this influence is unidirectional \cite{Lara-Cabrera2019-AffectiveGames}.
    \item Gamification: It refers to the application of game design concepts and elements in non-game contexts, with examples in recycling, health, learning, and work \cite{Chou2019-Gamification}.
    \item Exergames: Active games focused on physical activity, in which the player’s movements are synchronized with an avatar. This connection lasts only during gameplay \cite{Williams2020-PhysiccalActivity}.
\end{itemize}

In contrast, LifeSync-Games (Figures 2 and 3) aim to create reciprocal interactions between real life and virtual \mbox{environments} to support human \mbox{development}. These games allow players to temporarily transfer real-life \mbox{attributes} (e.g., from running, reading, or working) to influence in-game \mbox{mechanics}. Once the effect fades, players are encouraged to re-engage with real-world \mbox{activities} to continue developing cognitively, physically, emotionally, and socially.

As a framework, LifeSync-Games is adaptable to both new and existing titles, including \mbox{entertainment} games, serious games, affective games, and exergames. 


\section{The LifeSync-Games Framework}


\subsection{Conceptual and Theoretical Foundations}

\subsubsection{Video games}

Video games are a form of interactive \mbox{entertainment} in which users engage with virtual worlds across diverse genres and purposes, including \mbox{entertainment}, education, and rehabilitation \cite{Qaffas2020-VideogameGenres}. As of July 2022, the global industry exceeded USD\$200 billion and encompassed nearly 3 billion players \cite{Palma-Ruiz2022-GamingIndustry}. \mbox{Development} involves both major companies and independent developers — behind titles such as Among Us, Terraria, and Minecraft  — with growth fueled by modern game engines like Unity, Unreal Engine, Godot, and GameMaker \cite{Mohd2023-ModernGameEngines}, \cite{Vohera2021-DifferentGameEngines}.

\subsubsection{Responsible video game design and use}

Responsible video game design seeks to enhance benefits and reduce harms \cite{Criffiths2017-ProblematicVideogames}, while responsible use relies on players’ self-regulation \cite{Luxford2022-SelfRegulation}. However, many modern games prioritize commercial goals that encourage extended play \cite{Criffiths2017-ProblematicVideogames}, placing much of the regulatory responsibility on parents and institutions.

\subsubsection{Human \mbox{development}}

Human \mbox{development} involves \mbox{enhancing} physical, mental, and social well-being \cite{Chaaban2016-HumanDevelopment}. \mbox{Excessive} video game use can negatively impact physical health by promoting sedentary \mbox{behavior}, poor nutrition, fatigue, and obesity \cite{DeLisi2013-ViolentVideoGames}, as most games do not encourage physical activity beyond traditional controllers \cite{Turel2016-VideogameAddiction}.

In terms of mental well-being, regulated gameplay can foster positive emotions and resilience \cite{Johannes2021-VideogameWellbeing}, but \mbox{excessive} use may cause isolation, negative emotions, or addiction \cite{DeLisi2013-ViolentVideoGames}.

Socially, video games can broaden interactions; however, these are often superficial or disconnected from players’ real identities, potentially impeding the \mbox{development} of authentic relationships and values like empathy and tolerance \cite{Collins2013-ProblematicNonProblematic}.

\subsubsection{Digital twin}

In LifeSync-Games, a digital twin \mbox{approach} bridges the real and virtual worlds by digitally representing the player across multiple dimensions (e.g., physical, mental, social) \cite{Ferko2022-ArchitectingDT}.

Unlike industrial uses of digital twins for prediction and simulation, here the goal is to project real-world player \mbox{attributes} into the virtual environment using sensor data (see III.A.5. Sensors). These digital twin \mbox{attributes} are mapped onto the player’s in-game avatar \cite{Salagean2023-VirtualTwin}, integrating real traits such as strength, speed, attention, and empathy into gameplay. As these \mbox{attributes} are consumed during play, users are motivated to develop them in real life (Figures 2 and 3). Thus, avatars serve as reflective mirrors, with the potential to encourage positive \mbox{behavior}s and enhance players’ overall well-being.

\begin{figure}[htbp]
\centerline{\includegraphics[angle=0, width=9cm]{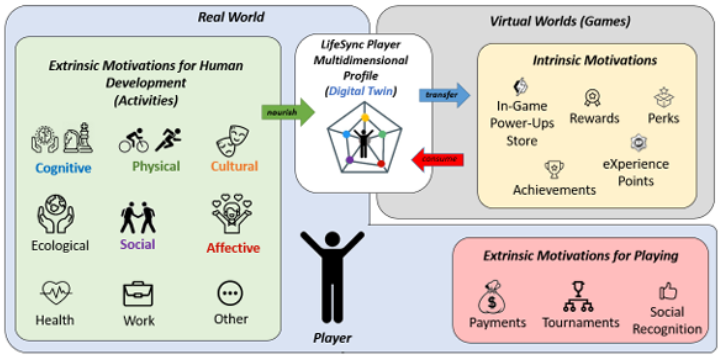}}
\caption{Conceptual model of LifeSync-Games with a simplified digital twin, linking the real and virtual worlds.}
\label{fig4-sccc}
\end{figure}

\begin{figure}[htbp]
\centerline{\includegraphics[angle=0, width=9cm]{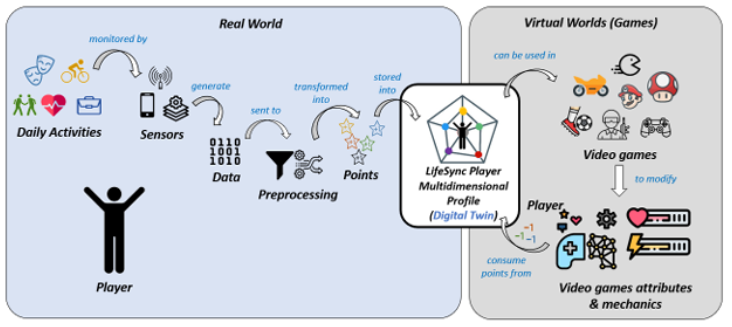}}
\caption{Conceptual model of the LifeSync-Games data flow, from the real world to virtual worlds.}
\label{fig3-sccc}
\end{figure}

\subsubsection{Sensors}

In LifeSync-Games, sensors are key to \mbox{bridging} the real and digital worlds. Physical sensors — such as \mbox{accelerometers}, GPS, and heart rate monitors \cite{Shin2024-EficienciaGemelosDigitales} — capture data from \mbox{players'} real-world actions, while virtual sensors extract and process information from existing digital sources, such as social media \cite{Hinduja2022-SocialsensorService}.

Both types create a multimodal sensory environment, \mbox{gathering} data from varied sources and modalities that \mbox{correspond} to the physical, mental, and social aspects of human \mbox{development}. This capability allows LifeSync-Games to effectively map and process complex signals to support the player’s digital twin.

\subsubsection{Emotion and Motivation in Video Games}

Emotions, shaped by motor, physiological, and cognitive factors, play a crucial role in influencing \mbox{behavior}. Constructivist psychology explains emotions as emerging from the interaction between social and psychological constructs and brain architecture \cite{Gendron2009-EmotionInPsychology}, \cite{Feldman2017-ComoSeContruyenLasEmociones}. This framework has guided research into basic emotions, including facial and \mbox{behavior}al expressions, and their relation to video games. For example, \mbox{studies} indicate that violent video games may impair the recognition of positive emotions \cite{Johannes2021-VideogameWellbeing}, while also fostering prosocial \mbox{behavior}s. Among adolescents, video games serve as a platform for emotional exploration and expression \cite{Farber2021-SocialAndEmotional}. Well-crafted video games enable players not only to express but also to explore and regulate emotions, providing safe \mbox{environments} for engaging with complex \mbox{feelings} and developing socio-emotional skills \cite{Löytömäki2023-SocialEmotionalskills}. Moreover, the emotional influence of video games extends beyond the digital space, shaping attitudes, empathy, and \mbox{behavior}s both inside and outside the game \cite{Gao2017-EmpathyandMorality}.

Emotions in video games cannot be fully understood through facial and \mbox{behavior}al cues alone, as their interpretation depends on the cultural and sociocultural context of players. Effective game design should incorporate these factors to better capture emotional complexity, thereby enriching the gaming experience and promoting social learning and well-being \cite{Toh2022-SocialEmotionalConcepts}.

Motivation, which refers to the biological, cognitive, and social factors that drive \mbox{behavior} \cite{Filipp1996-MotivationadnEmotion}, \cite{Reeve2018-UnderstandingMotivandEmot}, has been explained through \mbox{various} models, including those based on instincts, needs, incentives, and cognition. The widely accepted self-determination theory distinguishes intrinsic motivation — engaging in \mbox{activities} for inherent enjoyment — from extrinsic motivation, which depends on external rewards \cite{Ryan2000-IntrinsicAndExtrinsicMotiv}. Research indicates that extrinsic rewards may diminish intrinsic motivation in children. Furthermore, motivations for playing video games are diverse. For instance, in some cases, video games can support child \mbox{development} \cite{Chaarani2022-CognitivePerformanceAmongChild}, while multiplayer formats, such as dance games, can enhance player motivation \cite{Ferguson2013-MotivationForVideoGame}.

Recent research suggests rethinking motivation types: intrinsic motivation (inherent enjoyment) doesn’t always guarantee high performance, and extrinsic motivation (linked to rewards) is not always externally controlled. Achievement motivation, focused on goal pursuit, is emerging as a distinct category \cite{Locke2019-IntrinsicAndExtrinsicMotivation}. Understanding these distinctions can guide game design to integrate intrinsic, extrinsic, and achievement motivations, encouraging meaningful actions inside and beyond the virtual world.

These findings highlight the importance of incorporating diverse motivational factors in video game design to promote well-being. They also suggest that combining motivational and attentional strategies can enhance sustained player engagement \cite{Laine2020-GamesEducation} and adherence \cite{Dilla2009-AdherenciaTerapeutica}, supporting both in-game immersion and the achievement of real-life goals—especially in health and personal \mbox{development} contexts.

\subsubsection{Regulation}

Scientific \mbox{evidence} highlights both the risks associated with video games — such as sedentary \mbox{behavior}, sleep disruption, and poor self-regulation \cite{Lebby2023-AssociationBetweenVGandCG} — and their potential to encourage physical activity and positive change through thoughtful design. Although games frequently combine intrinsic and extrinsic motivations, the absence of \mbox{effective} self-regulation mechanisms can contribute to harmful habits \cite{Kökönyei2019-CognitiveEmotionRegulation}.

Self-regulation is a promising \mbox{approach} to foster players’ autonomy over their gaming habits, and video games must actively facilitate this process, signaling a paradigm shift toward designing experiences that promote balanced and mindful play.

In LifeSync-Games, self-regulation involves players \mbox{consciously} developing healthy habits and managing their gameplay in a balanced manner \cite{Luxford2022-SelfRegulation}. This is encouraged by \mbox{allowing} real-world actions to meaningfully influence gameplay and rewards (Figures 2 and 3).

Promoting player autonomy and responsibility is a desirable goal that requires thoughtful game design \cite{Ryan2006-SelfDetermination}. This design \mbox{approach} encourages moderate use by balancing escapism-driven motivations \cite{Kuo2016-ActiveEscapismWoW} with conditions that foster flow \cite{Csikszentmihalyi1975-BeyondBoredomAndAnxiety}, enabling gaming to serve not only as an escape but also as a tool for building healthy and sustainable habits.

\subsection{Technology}

LifeSync-Games aims to mitigate the negative effects of \mbox{excessive} video game exposure — such as addiction and sedentary \mbox{behavior} \cite{DeLisi2013-ViolentVideoGames} — by promoting a balanced interaction between digital activity and real life. To achieve this, it integrates real-world data, transforming it into \mbox{attributes} that influence game \mbox{mechanics} and enrich the player experience.

Its modular architecture enables efficient data acquisition, processing, storage, and use, resulting in a flexible and \mbox{scalable} technological ecosystem. The technological component is composed of the following six elements:

\subsubsection{Cloud module}

The cloud module is a core component of LifeSync-Games, responsible for storing and managing user profiles (i.e., \mbox{players'} digital twins), recording points derived from sensor data, and facilitating their integration into video games \cite{Calistro2019-bGames}, \cite{Mahu2020-ModuloCloud}, \cite{Zelada2023-RestauracionModuloCloud}. Its key functions include account management, sensor connection and administration, data processing (including attribute acquisition and normalization), and enabling point redemption within the game.

The architecture of the cloud module is structured into three main layers, each with specific responsibilities to facilitate the capture, processing, and use of real-world data in video games (Figure 4).

\begin{itemize}
    \item Communication Layer: Handles interaction with external sources, including data collection from web APIs, mobile devices, and connected games. It comprises microservices for web and mobile data capture, cloud-to-desktop communication, and \mbox{mechanics} authorization.
    \item Composition layer: Processes and standardizes the captured data.
    \item Data layer: Responsible for data storage and management.
\end{itemize}

These layers interact through well-defined interfaces. For example, a request to modify a game mechanic passes through the communication layer, is \mbox{validate}d by the composition layer, and then updates the data in the storage layer.

\begin{figure}[htbp]
\centerline{\includegraphics[angle=0, width=9cm]{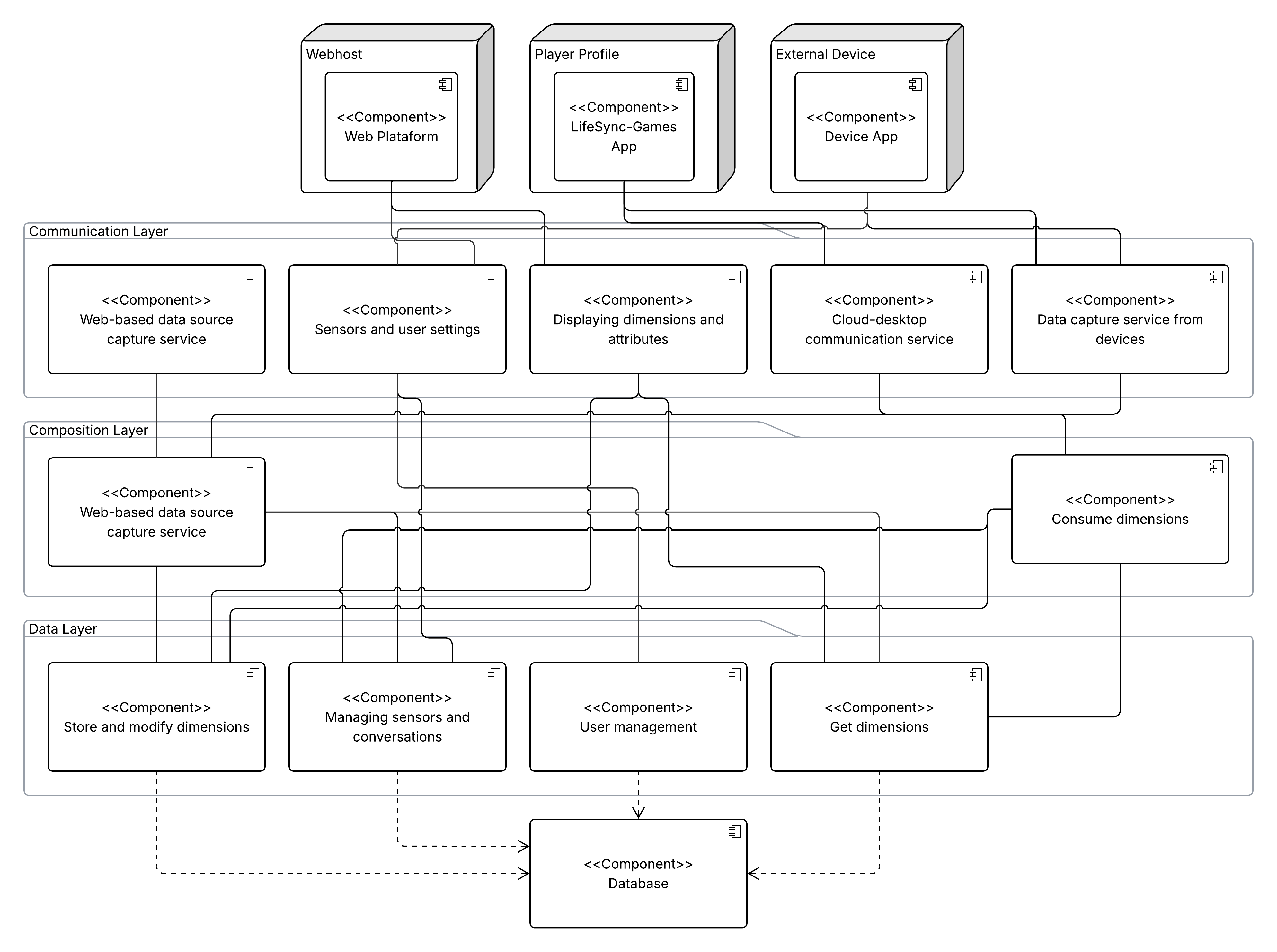}}
\caption{Components diagram of LifeSync-Games \cite{Mahu2020-ModuloCloud}.}
\label{fig2-sccc}
\end{figure}

\subsubsection{Application Programming Intefaces (APIs)}

APIs play a crucial role in LifeSync-Games by enabling developers to integrate the LifeSync-Games digital twin into video games.

More precisely, the cloud module communicates both internally and externally, primarily through RESTful APIs. These APIs support: (a) data capture from sensors, (b) point consumption within games, and (c) extensibility by enabling the integration of new data sources and features using standardized formats such as JSON.
\\

\subsubsection{Connectors}

Connectors are components that simplify the integration of LifeSync-Games into game engines and modding, enabling seamless implementation. 

Currently, connectors are \mbox{available} for Unity \cite{Ruz2022-JuegoRPG}, \cite{Piccinini2024-JuegoBulletLand}, GameMaker \cite{Lizama2022-JuegoStreetBlocks}, \cite{Ternero2022-JuegoVillageDefender}, Godot \cite{Fiorentino2024-JuegoNightmareSurvivors}, Unreal \cite{Onetto2023-JuegoBlazingDuel}, \cite{Fernandez2023-JuegoFPSSimulator}, Minecraft (Frabric and Forge) \cite{Simken2023-ModMinecraft}, Terraria \cite{Muñoz2024-ModTerraria}, Stardew Valley \cite{Godoy2024-ModStardewValley}, Cities: Skylines \cite{Aldea2025-ModCitiesSkylines}, and Starbound \cite{Herrera2025-ModStarbound}.
\\

\subsubsection{Sensors}

Sensors are essential in LifeSync Games, capturing real-world data and converting it into insights that enrich the user’s multidimensional profile. These sensors are based on either physical devices or software (i.e., virtual sensors).
\\

\textbf{Out-of-game sensors}

Given the focus on creating a reliable representation of the player in the real world, most LifeSync-Games sensors are designed to capture representative data from players’ real-world \mbox{activities} across diverse contexts.

For example, hardware sensors targeting the cognitive dimension include electroencephalography (EEG) headsets and eye trackers, which capture data on players’ attention during cognitively demanding tasks such as reading and working. In the physical dimension, pedometers, global positioning system (GPS) devices, heart rate monitors, and \mbox{accelerometers} provide valuable information about players’ endurance during \mbox{activities} like running or walking. For the affective dimension, specialized cameras and electrodermal activity (EDA) sensors are well suited to detect players’ emotional states and arousal levels.

Virtual sensors can target a wide range of \mbox{attributes} across the player’s multiple dimensions. For example, at the cognitive level, a sensor may capture data related to a player’s productivity in tasks such as coding, writing, project management, investing, and learning. Similarly, social sensors can collect data from players’ interactions with others beyond the realm of video games, including both face-to-face communication and activity on social media platforms \cite{Muñoz2024-ModTerraria}.
\\

\textbf{In-game sensors}

Although the primary focus is on \mbox{activities} outside the game, sensor integration also enables real-world data to directly influence game \mbox{mechanics}, creating an active feedback loop. For example, in Digital Masters \cite{Vargas2024-JuegoSerioDigitalMasters}, motion sensors (such as pedometers) link real-life physical activity to in-game actions, promoting healthy habits in an immersive way and making the game a reflection of real-world \mbox{behavior}. Similarly, a version of the classic Pong that uses EEG data to adjust the player’s paddle size according to their focus level during gameplay \cite{Calistro2019-bGames} was developed.

This type of sensor can also capture in-game data for research purposes. For example, data loggers can provide valuable insights into how players use their LifeSync-Games profiles during gameplay.
\\

\subsubsection{Game catalog}

LifeSync-Games has evolved from simple Unity demos to a diverse catalog of fully developed titles created in GameMaker, Unreal Engine, Godot, and Unity. The catalog includes original games spanning genres such as RPGs\cite{Ruz2022-JuegoRPG} and srategy \cite{Ternero2022-JuegoVillageDefender}. 
Additionally, LifeSync-Games have been applied to develop mods for popular titles such as Minecraft \cite{Simken2023-ModMinecraft} and Terraria \cite{Muñoz2024-ModTerraria}. More recently, the framework has been applied to serious games centered on financial education \cite{Soto2024-JuegoSerioWealthQuest} and entrepreneurship \cite{Vargas2024-JuegoSerioDigitalMasters}. All games and mods are open-source projects, contributing to a growing catalog that demonstrates the framework’s versatility and applicability across diverse game genres and platforms. Furthermore, as open-source initiatives, they provide valuable references and serve as successful examples for developers interested in applying LifeSync-Games into their own projects.

Table 1 presents a comprehensive list of the current LifeSync-Games catalog.
\\

\subsubsection{Sensor catalog}

The integration of sensors in LifeSync-Games enables the mapping of real-life \mbox{activities} into the game environment through a simplified digital twin or multidimensional profile (e.g., physical, mental, and social), managed by the cloud module \cite{Mahu2020-ModuloCloud}, \cite{Zelada2023-RestauracionModuloCloud}. \mbox{Various} sensors have been implemented to support this \mbox{approach}: cognitive sensors (e.g., focus and attention \cite{Calistro2019-bGames}, coding activity \cite{Riquelme2023-SensoresbGames}, financial habits \cite{Soto2024-JuegoSerioWealthQuest}, project management in Trello \cite{Mahu2020-ModuloCloud}, chess playing \cite{Mahu2020-ModuloCloud}, general productivity \cite{Muñoz2024-ModTerraria}); social sensors (e.g., activity on X [formerly Twitter], StackOverFlow \cite{Godoy2024-ModStardewValley}, Reddit \cite{Godoy2024-ModStardewValley}, Twitch \cite{Herrera2025-ModStarbound} and Discord \cite{Piccinini2024-JuegoBulletLand}, \cite{Aldea2025-ModCitiesSkylines}); physical sensors (e.g., heart rate monitors, pedometers \cite{Mahu2020-ModuloCloud}, \cite{Vargas2024-JuegoSerioDigitalMasters}); and healthy \mbox{behavior} sensors (e.g., screen time exposure \cite{Fiorentino2024-JuegoNightmareSurvivors}, and duration of gameplay \cite{Godoy2024-ModStardewValley}). This flexible architecture positions LifeSync-Games as an extensible and adaptable platform for research and a variety of game genres.

\subsubsection{Showcase Platform}

As a new paradigm in video games, LifeSync-Games faces critical \mbox{challenges} related to adoption and the formation of a sustainable community. These \mbox{challenges} involve not only players and game developers but also the broader gaming industry, as LifeSync-Games questions one of its core pillars: traditional engagement strategies.

To address these \mbox{challenges}, several approaches have been implemented. First, the game catalog has been diversified across genres, platforms, and game engines to meet the needs of a heterogeneous community. Second, sensor coverage has been expanded to capture a wide range of players’ real-world \mbox{attributes}. Third, popular titles have been modded to help reach larger audiences. Finally, the LifeSync-Games Showcase — a web platform built around the social regulation of content — was created to support a growing community. On this platform, developers can publish their games via GitHub repositories, and end-users can access and engage with these titles \cite{Campos2021-DashboardbGames}.  

\begin{table*}[htbp]
\caption{List of Video Games and Mods in the LifeSync-Game Catalog}
\scriptsize
\centering
\begin{tabular}{ c|c|c|c|c|c }
\hline
\textbf{Authors} & \textbf{Title} & \textbf{Genre} & \textbf{Game Engine} & \textbf{Type} & \textbf{Platform} \\
\hline
    \cite{Ruz2022-JuegoRPG} & Spirit Adventure & RPG & Unity & Game & Windows/Linux \\
    \cite{Ternero2022-JuegoVillageDefender} & Village Defender & Strategy & Game Maker & Game & Windows/Linux \\
    \cite{Lizama2022-JuegoStreetBlocks} & Street Blocks & Beat'em Up & Game Maker 2 & Game & Windows/Linux \\
    \cite{Onetto2023-JuegoBlazingDuel} & Blazing Duel & Fighting & Unreal & Game & Windows \\
    \cite{Fernandez2023-JuegoFPSSimulator} & Zona Cero & Shooter & Unreal & Game & Windows \\
    \cite{Fiorentino2024-JuegoNightmareSurvivors} & Nightmare Survivor & Survivors & Godot & Game & Windows \\
    \cite{Piccinini2024-JuegoBulletLand} & Bulletland & Roguelike/Bullet hell & Unity & Game & Windows \\
    \cite{Soto2024-JuegoSerioWealthQuest} & WealthQuest & Educational & Unity & Serious Game & Windows/Linux \\
    \cite{Vargas2024-JuegoSerioDigitalMasters} & Digital Masters & Educational & Unity & Serious Game & Android \\
    \cite{Simken2023-ModMinecraft} & Minecraft & Sandbox & Java & Mod & Windows/Linux/Android \\
    \cite{Godoy2024-ModStardewValley} & Terraria & Sandbox & C\# & Mod & Windows \\
    \cite{Muñoz2024-ModTerraria} & Stardew Valley & RPG & C\# & Mod & Windows \\
    \cite{Herrera2025-ModStarbound} & Starbound & Sandbox & LUA & Mod & Windows \\
    \cite{Aldea2025-ModCitiesSkylines} & Cities: Skylines & Simulator & C\# & Mod & Windows \\
\hline
\end{tabular}
\label{tab:game_catalog}
\end{table*}

\subsection{Evaluation}

Evaluation within LifeSync-Games is a continuous and multifaceted process, crucial for validating the framework’s effectiveness, its integration into video games, and its impact on human development. The process emphasizes technical quality, user experience, and requirement compliance, laying the groundwork for future research into user behavior.

So far, the evaluation of LifeSync-Games has followed a pragmatic and progressive \mbox{approach}, prioritizing early validation and adaptation to limited resources. This includes the following approaches:

\begin{itemize}
    \item Prototyping for validation: Multiple prototypes are developed to \mbox{validate} game \mbox{mechanics}, refine ideas, and gather design feedback prior to full implementation. A recommended \mbox{approach} used in ongoing LifeSync-Games applications is a simplified version of Rapid Application \mbox{development} (RAD) \cite{Martin1991-RAD}.
    \item Testing with trusted users: Initial testing was conducted with InTeractiOn-Lab members to gather early feedback before broader empirical \mbox{studies}. Evaluation methods may include standardized questionnaires (e.g., \cite{Desurvire2004-HEP},\cite{Peres2013-SUS}) as well as qualitative techniques such as interviews and think-aloud protocols.
    \item Rigorous technical evaluation: Evaluation involves \mbox{various} forms of testing, including performance (CPU, GPU, RAM, FPS, latency), software quality (e.g., coverage, unit and integrion tests), platform compatibility, and data integrity. For mods, it is particularly important to assess the performance impact of integrating LifeSync-Games into existing titles \cite{Simken2023-ModMinecraft}. Gathered data is then analyzed using quantitative methods.
\end{itemize}

Although preliminary, these approaches offer valuable \mbox{evidence} for the ongoing improvement of the framework and lay the groundwork for future research into its impact on responsible gaming and human \mbox{development}.

However, despite being resource-efficient, the main \mbox{challenge} in evaluation lies in validating the core hypotheses of LifeSync-Games — specifically, their intended impact on human \mbox{development}. \mbox{addressing} this \mbox{challenge} requires mixed-methods research conducted through longitudinal \mbox{studies} aimed at investigating both the short- and long-term effects of these games on physical, mental, and social well-being, as well as on self-regulation.

\subsection{General Guidelines for Applying LifeSync-Games}

The components of the framework described above are designed to provide conceptual, theoretical, and technological resources for both game developers and researchers. To apply these components in new or existing games, designers and developers can follow the steps outlined below.   

\begin{enumerate}
    \item Game design document (GDD): Whether working on a new or existing title, game designers should begin by specifying game \mbox{mechanics} and other relevant elements in a formal GDD \cite{Adams2006-FundamentalsGameDesign}. It is essential to design the game to function both with and without LifeSync-Games integration. This ensures that players with or without a LifeSync-Games profile can enjoy the game, making the account linkage optional for those who have one.
    
   Once the game \mbox{mechanics} are clearly defined, designers must determine which ones will be influenced by \mbox{players'} real-world \mbox{attributes} and in what ways. They may opt to incorporate both positive and negative \mbox{attributes} to shape gameplay. For instance, if a player exhibits low physical activity in real life, their in-game avatar might move more slowly. Conversely, a physically active player might benefit from enhanced speed or agility during gameplay.
    
    Additionally, designers should determine whether these \mbox{attributes} will be dynamically integrated during gameplay (e.g., automatically mapped to game \mbox{mechanics}) or triggered on demand (e.g., used as power-ups).
    
    \item Game prototyping: Developers build initial game prototypes based on the GDD, integrating game \mbox{mechanics} with players’ multidimensional \mbox{attributes} via the LifeSync-Games digital twin. The aim is to test and \mbox{validate} how real-world data shapes the virtual experience. Prototypes can use either a local or public deployment of the LifeSync-Games cloud module. A local setup allows direct modification of player \mbox{attributes}, streamlining and accelerating testing without sensor reliance. The GDD may be revised as needed before moving to the next phase.
    
    \item Game \mbox{development}: Next, the process continues with the full \mbox{development} of the video game, building upon the \mbox{validate}d prototypes. This phase involves architectural design and the actual implementation of the game according to the GDD. An iterative and incremental \mbox{development} \mbox{approach} is recommended. Developers may initially use a local installation of the LifeSync-Games cloud module; however, transitioning to a publicly \mbox{available} version is necessary to conduct formal testing.   
    
    \item Game Evaluation: At this stage, the quality and \mbox{playability} of the video game must be evaluated \mbox{using} both quantitative and qualitative methods. First, a technical evaluation should be conducted to ensure performance, software quality, and seamless integration with LifeSync-Games. Second, user studies should be designed and carried out to assess the overall gameplay experience, incorporating instruments such as Heuristic Evaluation for Playability (HEP) \cite{Desurvire2004-HEP} to systematically identify usability and \mbox{playability} issues.
    
    \item Game publication: Finally, this phase focuses on \mbox{releasing} the video game — along with its components for integrating multidimensional \mbox{attributes} — to the broader community of users, developers, and researchers. Distribution channels may include the LifeSync-Games showcase platform, public repositories, and specialized modding or game distribution sites.
\end{enumerate}


\section{Exploratory applications of the framework}


Regarding mod \mbox{development}, this has served as a dissemination strategy that has proven \mbox{effective}, though with a moderate impact in reaching a broader end-user audience.  

A list of audiovisual material can be found on our official Website \cite{LSG-OfficialWebsite} and in our public YouTube channel \cite{LSG-YoutubeChannel}.

In the traditional video game model, a player’s character progresses and improves primarily through in-game actions (Figure 1). In contrast, the LifeSync-Games model (Figures 2 and 3) connects real-life \mbox{activities} — such as physical exercise, work, or social interaction — to game \mbox{mechanics}. This establishes a reciprocal relationship in which in-game performance and progression may depend on the player’s real-world \mbox{behavior}, aiming to encourage healthy habits and promote human \mbox{development}.

To date, a diverse catalog of video games and mods developed in accordance with the LifeSync-Games guidelines has been produced. This collection includes prototypes, full games, and mods for popular titles (Table 1), all of which are open-source projects aimed at promoting developer engagement and adoption of the framework. Technical evaluations indicate strong performance across the board, with minimal to no impact on system resources (e.g., CPU, GPU, FPS), along with positive feedback from players. These projects demonstrate the framework’s versatility across a wide range of genres, game engines, programming languages, and platforms. Moreover, they reflect the early adoption of LifeSync-Games by a growing community of developers. 


\section{Conclusions}


LifeSync-Games represents a significant advancement in integrating everyday life with digital gaming, fostering a balance between \mbox{entertainment} and personal well-being. By leveraging multidimensional profiles — represented as simplified digital twins and powered by data from real-world \mbox{activities} — game \mbox{mechanics} can be adapted across diverse genres, platforms, and game engines.

Despite these achievements, \mbox{development} has encountered \mbox{challenges}, and several key areas still need to be strengthened and consolidated to advance this new paradigm.

\subsection{Key \mbox{challenges} and Research Opportunities}

The implementation and adoption of LifeSync-Games have encountered several \mbox{challenges} inherent to its innovative nature and the complexity of the game \mbox{development} ecosystem:

\begin{itemize}
    \item Visibility and adoption: It has yet to achieve the \mbox{anticipated} reception within the community, making it \mbox{challenging} to establish a solid user and developer base.
    \item Limited initial diversity: At first, the limited catalog of games and sensors constrained the perception of its full potential.
    \item Technical \mbox{challenges} of the cloud module: \mbox{challenges} related to obsolescence and resource loss have affected \mbox{development} and distribution, necessitating restoration and continuous maintenance efforts.
    \item Integration and latency: Ensuring smooth, low-latency communication between sensors, the cloud module, and video games remains a \mbox{challenge}, particularly in multiplatform \mbox{environments}.
    \item Legal and ethical aspects: Capturing and storing sensitive data raises critical privacy, security, and ethical concerns that require thorough and proactive mitigation.
    \item Balance and user experience: Designing gameplay that is both engaging and well-balanced — through mechanics, design, and playability — while supporting health or educational goals remains a persistent \mbox{challenge}.
    \item Artificial intelligence (AI) integration and reliability: LifeSync-Games can benefit from AI both by \mbox{enhancing} the gaming experience and by analyzing interaction \mbox{patterns} between the real and virtual worlds.
    \item Sensors reliability: Although sensors are built to capture real-world activity data, they face significant \mbox{challenges} from manipulation and cheating. AI can help preserve system integrity by detecting fraud or misuse through the identification of anomalous or inconsistent behavior.
\end{itemize}

\subsection{Future work}

The extensible potential of LifeSync-Games opens multiple pathways to consolidate and expand its impact:

\begin{itemize}
    \item Expansion of sensors and data sources: Integrate additional physical sensors, online data sources, and advanced analytics modules to create more sophisticated user profiles; explore metrics derived directly from video games; and enhance security and authentication measures.
    \item Expansion to new platforms and game engines: Continue \mbox{development} for untargeted systems and consoles, and develop standardized tools and documentation to streamline integration for developers.
    \item Impact evaluation: Conduct longitudinal and collaborative \mbox{studies} on sedentary \mbox{behavior}, addiction, healthy habits, and educational outcomes.
    \item Framework and content improvements: Refine code, deepen player profiling and analytics, improve mobile interface, enhance game balance and feedback, and enrich content and narrative.
    \item \mbox{Attributes} standardization: Develop a formula to quantify the value of captured data, balancing the multidimensional aspects of the player profile with in-game rewards.
\end{itemize}

\mbox{Addressing} these \mbox{challenges} will position LifeSync-Games as both a technological framework and a design paradigm, promoting conscious interaction between real and virtual worlds to support self-regulation and human \mbox{development}.


\section{Acknowledgments}


We thank the undergraduate students who contributed to the \mbox{development} and validation of the LifeSync-Games framework, and acknowledge the use of ChatGPT v4.0o and Google NotebookLM for manuscript proofreading.

\vspace{12pt}

\end{document}